\documentclass[twocolumn,aps,superscriptaddress,nofootinbib,floatfix,linenumbers]{revtex4}
\usepackage{amsfonts}
\usepackage{float}

\usepackage{placeins}
\usepackage{graphicx}
\usepackage[hidelinks]{hyperref}
\usepackage{color,amsmath,amssymb,bm}
\usepackage{tensor}
\usepackage[cmyk]{xcolor}

\usepackage[normalem]{ulem}  

\renewcommand{\sout}{\bgroup \color{black} \ULdepth=-.5ex \ULset}

\def\blfootnote{\xdef\@thefnmark{}\@footnotetext}

\newcommand{\beq}{\begin{equation}}
	\newcommand{\eeq}{\end{equation}}
\newcommand{\bea}{\begin{eqnarray}}
	\newcommand{\eea}{\end{eqnarray}}

\newcommand{\gtsim}{\raisebox{-4pt}{$\,\stackrel{\textstyle >}{\sim}\,$}}

\begin{document}
	
	\title{
		B meson production in Pb+Pb at 5.02 ATeV at LHC: estimating the diffusion coefficient in the infinite mass limit
	}
	
	\author{Maria Lucia Sambataro}
	\email{mlsambataro@gmail.com}
	\affiliation{Department of Physics and Astronomy 'E. Majorana', University of Catania, Via S. Sofia 64, 1-95123 Catania, Italy}
	\affiliation{Laboratori Nazionali del Sud, INFN-LNS, Via S. Sofia 62, I-95123 Catania, Italy}
	
	\author{Vincenzo Minissale}
	\email{vincenzo.minissale@lns.infn.it}
	\affiliation{Laboratori Nazionali del Sud, INFN-LNS, Via S. Sofia 62, I-95123 Catania, Italy}
	
	\author{Salvatore Plumari}
	\email{salvatore.plumari@dfa.unict.it}
	\affiliation{Department of Physics and Astronomy 'E. Majorana', University of Catania, Via S. Sofia 64, 1-95123 Catania, Italy}
	\affiliation{Laboratori Nazionali del Sud, INFN-LNS, Via S. Sofia 62, I-95123 Catania, Italy}
	
	\author{Vincenzo Greco}
	\email{greco@lns.infn.it}
	\affiliation{Department of Physics and Astronomy 'E. Majorana', University of Catania, Via S. Sofia 64, 1-95123 Catania, Italy}
	\affiliation{Laboratori Nazionali del Sud, INFN-LNS, Via S. Sofia 62, I-95123 Catania, Italy}

	\date{\today}
	
	\begin{abstract}
		
		In the last decade a Quasi-Particle Model (QPM) has been developed to study charm quark dynamics in ultra-relativistic heavy-ion collisions supplying a satisfactory description of the main observables for $D$ meson and providing an estimate of the space-diffusion coefficient $D_s(T)$ from the phenomenology.
		In this paper, we extend the approach to bottom quarks describing their propagation in the quark-gluon plasma within an event-by-event full Boltzmann transport approach followed by a coalescence plus fragmentation hadronization. 
		We find that QPM approach is able to correctly predict the first available data on $R_{AA}(p_T)$ and $v_{2}(p_T)$ of single-electron from B decays without any parameter modification w.r.t. the charm. We show also predictions for centralities where data are not yet available for both $v_{2}(p_T)$ and $v_{3}(p_T)$.
		Moreover, we discuss the significant breaking of the expected scaling of the thermalization time $\tau_{th}$ with $M_Q/T$, discussing the evolution with mass of $D_s(T)$ to better assess the comparison to lQCD calculations. We find that at $T=T_c$ charm quark $D_s(T)$ is about a factor of 2 larger than the
		asymptotic value for $M \rightarrow \infty$, while bottom $D_s(T)$ is only a $20-25\%$ higher. This
		implies a $D_{s}(T)$ which is consistent within the current uncertainty to the most recent lattice QCD calculations with dynamical quarks for $M \rightarrow \infty$. 
	\end{abstract}
	
	\maketitle
	
	\section{Introduction}
	
	The main goal of the ongoing heavy-ion
	collisions performed at Relativistic Heavy Ion
	Collider (RHIC) and Large Hadron Collider (LHC)
	is the study of a state of matter named Quark-Gluon Plasma(QGP) that behaves like a nearly perfect fluid having a remarkably small value of shear viscosity to entropy density ratio, $\eta/s \approx 0.1$. Heavy quarks,
	namely charm and bottom, thanks to their large
	masses, are considered as a solid probe to characterize the QGP phase \cite{Dong:2019unq,He:2022ywp,Andronic:2015wma}. They are produced by pQCD processes, hence at variance with the bulk matter, their initial production is to a large extent known. Furthermore, they have a formation time $\tau_{0}<0.08 fm/c\ll\tau_{QGP}$ so probing also the strong electromagnetic fields expected in the initial stage of the collision \cite{Das:2016cwd, Jiang:2022uoe, Chatterjee:2018lsx}.  The large mass implies a larger thermalization time w.r.t. light counterpart and appears currently to be comparable to the one of the QGP itself \cite{Dong:2019unq,Scardina:2017ipo}. Therefore, HQs can probe the whole evolution of the plasma and, being produced out-of-equilibrium, they are expected to conserve memory of the history of the system evolution. 
	Furthermore, recently it has been suggested a relevance of the early glasma phase on their
	dynamics in both pA and AA collisions \cite{Sun:2019fud,Liu:2020cpj,Avramescu:2023qvv}. There is a general agreement that the observed $R_{AA}(p_T)$ and $v_2(p_T)$ imply that in the low-intermediate $p_T$ region charm quark dynamics is affected by large non perturbative effect \cite{vanHees:2005wb, vanHees:2007me, Gossiaux:2009mk, Dong:2019unq, Prino:2016cni, Berrehrah:2013mua,Cao:2015hia}. 
	In order to take into account the
	non-perturbative effects of the interactions, some approaches make use of pQCD framework \cite{Uphoff:2010sh,Uphoff:2011ad,Uphoff:2012gb} with large coupling,
	or supplemented by Hard-Thermal Loop (HTL) \cite{Gossiaux:2008jv,Alberico:2011zy}. Another 
	way to account the non-perturbative QCD effects at non-zero temperature
	is to encode the lQCD thermodynamical expectations with effective temperature
	dependent particle masses like in Quasi-Particle Model
	(QPM) \cite{Plumari:2011mk, Song:2015ykw, PhysRevC.96.014905} or similarly, but
	including the off-shell dynamics, in the DQPM \cite{Berrehrah:2013mua,Berrehrah:2014kba}.
	A more sophisticated approach is based on a T-matrix calculation under a potential kernel that correctly reproduce the free energy as evaluated in lattice QCD for HQ pair in the infinite mass limit
	\cite{vanHees:2007me,Liu:2018syc}.
	In the past years, the QPM approach has successfully described the main observables of $D$ mesons leading to an extrapolation of the spatial diffusion coefficient $D_s(T)$ of charm quark in agreement with the available lQCD calculation in quenched approximation \cite{Dong:2019unq,He:2022ywp,Scardina:2017ipo,vanHees:2005wb,vanHees:2007me,Gossiaux:2008jv,Das:2009vy,Alberico:2011zy,Uphoff:2012gb,Lang:2012nqy,Song:2015sfa,Song:2015ykw,Das:2013kea,Cao:2015hia,Das:2015ana,Cao:2017hhk,Das:2017dsh,Sun:2019fud,Cao:2018ews,Rapp:2018qla}.
	On the other hand, one has to consider that a proper comparison should be done with more recent calculation in lQCD where the quenched approximation is relaxed. Furthermore, the comparison with charm quark suffers from its finite mass that while being significantly larger than the QGP temperature is still nearly comparable with the 
	average thermal momentum $\sim 3T$ that is also comparable with the exchanged momentum
	$\sim g T$. In this respect, the extension of the study to the bottom sector allows to
	investigate the quark mass dependence of the interaction toward the infinite mass limit assumed in the present lQCD calculations 
	\cite{Banerjee:2011ra, Kaczmarek:2014jga, Francis:2015daa, Brambilla:2020siz, Altenkort:2023oms}. Hence from the phenomenological point of view, bottom allows also to test the scaling of the thermalization time $\tau_{th}$ with the heavy quark mass, an aspect that to our knowledge, we are focusing for the first time. 
	Two main observables have been studied in uRHICs for HF hadrons: the heavy mesons nuclear modification factor $R_{AA}(p_T)$ \cite{STAR:2006btx, PHENIX:2005nhb, ALICE:2015vxz} , and the so called elliptic flow,
	$v_2(p_T)$ \cite{PHENIX:2006iih, Abelev:2014dsa}. The first observable describes the change of the spectrum in nucleus-nucleus collision with respect to a simple proton-proton superposition, while the second is related to the anisotropy in the in the particle angular distribution giving information about the coupling of the HQs with the plasma. 
	Further efforts have been done to extend the analysis to higher order anisotropic flows $v_n$ \cite{Beraudo:2021ont, Katz:2019fkc, Ke:2018tsh, Sambataro:2022sns} that can give more constraints on the extraction of the transport coefficients which are strictly related to the initial event-by-event fluctuations. 
	Further investigation on the HQs dynamics can be addressed by using the Event-Shape-Engineering technique \cite{Schukraft:2012ah, ALICE:2020iug} that seem to be satisfactory described by models available in literature \cite{Beraudo:2018tpr, Sambataro:2022sns, PhysRevC.96.064903, Plumari:2019hzp}, at least within the current experimental data uncertainties.
	In this paper, within our approach already widely employed to study the charm dynamics, we want to study the bottom dynamics through the nuclear modification factor $R_{AA}(p_T)$, elliptic and triangular flows $v_{2,3}(p_T)$ of B mesons and electrons from semi-leptonic B meson decay which can be compared to the available experimental data from ALICE collaboration.
	Moreover, we discuss the extrapolation of the spatial diffusion coefficient $D_s$ for bottom quark, comparing our results with the $D_s$ of charm quark from the previous analysis and the available lQCD data points evaluated in the infinite mass limit for the heavy quarks.
	In particular, we discuss the $D_s$ dependence on the heavy quark masses in our QPM approach evaluating the discrepancy between both charm and bottom mass scale with respect to the saturation value of $D_s$ reached in the infinite mass limit. The paper is organized as follows. In section II, we expose briefly the Boltzmann transport approach used to
	describe the HQs evolution and the hybrid hadronization approach by coalescence plus fragmentation. In section III, the results for the main observables in the bottom sector are shown, in particular our predictions for the $R_{AA}$, $v_{2,3}$ of B mesons and electrons from semi-leptonic B meson decays. In the section IV, we discuss the $D_s(T)$ of bottom quark in comparison with the $D_s(T)$ obtained of charm quark from the previous analysis and the available lQCD data points. Finally, section V contains a summary and some
	concluding remarks.

	\section{Transport evolution of bottom quark in QGP}

	The results shown in this paper have been obtained using a transport code developed to perform studies of the dynamics of relativistic heavy-ion collisions at both RHIC and LHC energies \cite{Plumari:2012ep,Ruggieri:2013ova,Scardina:2014gxa,Plumari:2015cfa,Scardina:2017ipo,Plumari:2019gwq,Sun:2019gxg}.
	In our approach, the space-time evolution of gluons ($g$) and light quarks ($q$) as well as of heavy quarks ($Q$) distribution functions is described by mean of the Relativistic Boltzmann Transport (RBT) equations given by:
	\begin{eqnarray}
		& & p^{\mu}_i \partial_{\mu}f_{i}(x,p)= {\cal C}[f_q,f_g](x_i,p_i) \label{eq:RBTeqq} \\
		& & p^{\mu} \partial_{\mu}f_{Q}(x,p)= {\cal C}[f_q,f_g,f_{Q}](x,p) \label{eq:RBTeqQ}
		\label{B_E} 
	\end{eqnarray}
	where $f_i(x,p)$ is the on-shell phase space one-body distribution function for $i-th$ parton species ($i=q,g$) and ${\cal{C}}[f_q, f_g, f_{Q}](x,p)$ in the right-hand side of Eq. \ref{eq:RBTeqQ} is the relativistic Boltzmann collision integral allowing to describe the short range interaction between
	heavy quark and particles of plasma. In our calculations the HQs interact with the medium by
	mean of two-body collisions regulated by the scattering matrix of the processes $g + Q \to g + Q$ and $q(\bar q) + Q \to q(\bar q) + Q$ and the collision integral describing HQ scattering takes the form:
	
	\begin{align}\label{int_finale}
		\begin{split}
			C[f]=&\frac{1}{2E_1}\int\frac{d^3p_{2}}{2E_2 (2\pi)^3}\int\frac{d^3p'_1}{2E_{1\prime}(2\pi)^3}
			\\&\times[f_Q(p'_1)f_{g,q}(p'_2)-f_Q(p_1)f_{g,q}(p_2)]
			\\&\times{|\cal M}_{(g,q)+Q}(p_1p_2\rightarrow p'_1p'_2)|^2
			\\&\times (2\pi)^4\delta^4(p_1+p_2-p'_1-p'_2)
		\end{split}
	\end{align}  
	
	where $|{\cal M}_{(g,q)+Q}|^2$ are the transition amplitude of the process. As shown by the above eq.s, we are discarding the impact of heavy quarks (charm or bottom) on the bulk dynamics, which is quite a solid approximation.
	Furthermore, in our simulations we are employing a bulk with thermal massive quarks and gluons according to a Quasi-Particle Model (QPM) which is able to reproduce the lattice QCD Equation of State: pressure, energy density and interaction measure $T_{\mu}^{\mu}=\epsilon - 3P$, giving a softening of the equation of state  consistent with a decreasing speed of sound approaching the cross-over region \cite{Borsanyi:2010cj}.
	However, the main feature of the QPM on the HQ dynamics results in a significantly stronger coupling of HQs with the bulk medium respect to the pQCD coupling at lower temperature particularly as $T \rightarrow T_c$ (see details in Refs.~\cite{Das:2015ana,Scardina:2017ipo, Sambataro:2020pge}).
	In the collision integral $C[f_q, f_g](x,p)$ for gluon and light quark, the total cross section is determined in order to keep the ratio $\eta/s=1/(4\pi)$  fixed during the evolution of the QGP, see Refs.~\cite{Plumari:2019gwq,Plumari:2015cfa,Ruggieri:2013ova} for more details. In this way, we simulate the dynamical evolution of a fluid with specified $\eta/s$ by means of the Boltzmann equation. We include initial state fluctuations by means of a modified Monte Carlo Glauber model as used in Ref.  \cite{Sun:2019gxg} to study the light flavour $v_n$ and recently  extended to study the dynamics of charm quarks \cite{Sambataro:2022sns}. Charm and bottom quark r-space distributions follow the number of binary nucleon-nucleon collisions $N_{coll}$ from the Monte Carlo Glauber model.
	Further details of the initial condition implementation can be found in  \cite{Sun:2019gxg, Sambataro:2022sns}. At last, for the charm and bottom quark initial distributions in momentum space, we have used the spectra calculated at Fixed Order + Next-to-Leading Log (FONLL)~\cite{Cacciari:2012ny} which describe the D-meson spectra in proton-proton collisions after fragmentation. 
	
	In our simulations, the hadronization hypersurface is given by the space-time where the local temperature of a cell falls down the critical temperature $T=155\, \rm MeV$ in agreement with the statistical hadronization model \cite{Andronic:2019wva}.
	The corresponding distribution functions are employed to undergo the hadronization process by coalescence
	plus fragmentation; an approach that has been widely discussed and employed  for charm quark, for details see Ref.s \cite{Greco:2003vf,Plumari:2017ntm,Minissale:2019gbf}. In the following, we describe the main features and parameters for the bottom case. For the case of bottom quark, following Refs. \cite{Greco:2003vf,Oh:2009zj,Plumari:2017ntm,Minissale:2020bif} as for the charm quarks, we adopt as B mesons Wigner function a Gaussian shape in relative coordinates:
	\begin{equation}
		f_M(r,p)= 8 \exp{\Big(-\frac{x_{r}^2}{\sigma_{r}^2} - p_{r}^2 \sigma_{r}^2\Big)}
		\label{Eq:Wigner_MB}
	\end{equation}

	where $x_{r}=x_{1} - x_{2}$ and $ p_{r}=\frac{m_{2} p_{1}- m_{1} p_{2}}{m_{1}+m_{2}}$. The $\sigma_{r}$ are the widths which can be related to the root mean square charge radius of the hadron:
	\begin{eqnarray} 
		\langle r^2\rangle_{ch}&=& \frac{3}{2}  \frac{Q_1 m_2^2+Q_2 m_1^2}{(m_1+m_2)^2} \sigma_r^{2}
	\end{eqnarray}
	with $Q_i$ the charge of the i-th quark. 
	The width parameter $\sigma_{r}$ depends on the hadron species and can be calculated from the charge radius of the hadrons that have been taken from quark model \cite{Hwang:2001th,Albertus:2003sx} and the corresponding widths for B mesons are shown in Table \ref{table:param}.
	\begin{table} [ht]
		\begin{center}
			\begin{tabular}{l |c c c }
				\hline
				\hline
				Meson &$\langle r^2\rangle_{ch}$ & $\sigma_{p1}$ \\ 
				$B^{-}=[b \bar{d}]$     & 0.378   & 0.302 \\
				$B_{s}=[\bar{s}b]$   & -0.119   & 0.368  \\ 
				\hline
				\hline
			\end{tabular}
		\end{center}
		\caption{Mean square charge radius $\langle r^2\rangle_{ch}$ in $fm^2$ and the widths parameters $\sigma_{pi}$ in $GeV$. The mean square charge radius are taken quark model \cite{Hwang:2001th,Albertus:2003sx}.}
		\label{table:param}
	\end{table}
	The $B^\star$ resonant states are suppressed according to the statistical thermal weight with respect to the ground state. We consider the  $B^{*} (5325)=\bar{l}b$, 
	$B_1(5721)^{+0} =\bar{l}b$, $B_2(5747)^{+0} =\bar{l}b$, $\bar{B}_{s}^{*} =\bar{s}b$,
	$B_{s1}(5830)^{0} =\bar{s}b$, $B_{s2}(5840)^{0} =\bar{s}b$.
	Finally, as for the charm quark sector \cite{Plumari:2017ntm,Minissale:2020bif}, an overall normalization of the coalescence probability is fixed to guarantee
	that in the limit $p \to 0$ all the bottom quarks hadronize by coalescence in a heavy hadron. This is imposed by requiring that the total coalescence probability gives $\lim_{p \to 0} P^{tot}_{coal}(p)=1$.
	In our hybrid hadronization approach, the bottom quarks that do not hadronize via coalescence are converted to hadrons via fragmentation; with probability for each bottom quark given by $P_{frag}(p_T)=1-P_{coal}(p_T)$. 
	Therefore, in order to obtain the final hadron spectra coming from fragmentation; we evaluate the convolution, integrating over all the momentum fraction $z$, between the momentum distribution of heavy quarks which do not coalescence and the Kartvelishvili fragmentation function\cite{Kartvelishvili:1977pi} as implemented in FONLL:
	
	\begin{equation}\label{fragm}
		D(z) \propto z^ {\alpha}(1-z)
	\end{equation}

	where $z = p_{had}/p_b$ is the momentum fraction carried by the
	heavy hadron formed from the heavy quark fragmentation and
	$\alpha$ is a parameter that we determine to reproduce the experimental HF mesons in $pp$ collisions measured at LHC; in particular we obtain a value $\alpha=25$.
	
	\section{Nuclear modification factor $R_{AA}$ and anisotropic flows $v_{2,3}$ in bottom sector}
	We have generated our prediction for B meson observables employing the QPM modeling for the HQ interaction already employed to study the D meson dynamics \cite{Scardina:2017ipo,Plumari:2019hzp, Sambataro:2022sns}, described in Sect. II.
	We stress that the model parameters and in particular the coupling entering the HQ scattering matrices, have not been modified going from charm to bottom, hence the difference
	come merely from the different mass value.
	In our model, event-by-event fluctuations generate an initial profile in the transverse plane $\rho_\perp(\textbf{x}_\perp)$, that changes in every event, and which is responsible for the initial anisotropy in coordinate space. This anisotropy is quantified in terms of eccentricities $\epsilon_n$ of the initial fireball. The charm or bottom quarks interact with the QGP constituents during the evolution; they convert the initial eccentricity of the overlap region into a final anisotropy in momentum space, that is characterised in terms of the Fourier coefficients $v_n(p_T)$.The results shown in this paper have been obtained using the two particle correlation method to calculate elliptic ($v_2$) and triangular flow ($v_3$) \cite{Chatrchyan:2011eka,Chatrchyan:2012wg}.
	In order to guarantee a numerical solution that reaches convergence and stability for $R_{AA}(p_T)$ and $v_{2,3}(p_T)$ up to $p_T \sim \,$10 GeV$/c$, we have used a total number of test particles $N_{\rm{test}} = 4\cdot 10^5$ per unit of rapidity and a lattice discretization with $\Delta x=\Delta y=0.5$ fm and $\Delta \eta=0.1$. \\
	The only available observable to infer information about B mesons production in nucleus-nucleus collisions are the nuclear modification factor and elliptic flow of leptons coming from semi-leptonic B mesons decay in $Pb-Pb$ collisions at $\sqrt{s}_{NN}=5.02 \, \rm TeV$ at various centralities.
	In this section, we first show the comparison between our results and the experimental data for the nuclear modification factor $R_{AA}$ of electrons from semi-leptonic B meson decay and we provide predictions for B mesons $R_{AA}$ at LHC energies for two different centrality classes. The term "electron" throughout this paper is used for indicating both electrons and positrons.
	We evaluate the nuclear modification factor $R_{AA}(p_T)$ as the ratio between the particle spectrum in nucleus-nucleus collisions and the spectrum in proton-proton collisions scaled with the number of binary collisions. The modification of the parton distributions in nuclei, referred to as shadowing, has been taken into account by the parametrization provided in EPS09 \cite{Eskola:2009uj}.
	\begin{figure}[h]
		\begin{center}
			\includegraphics[width=.45\textwidth]{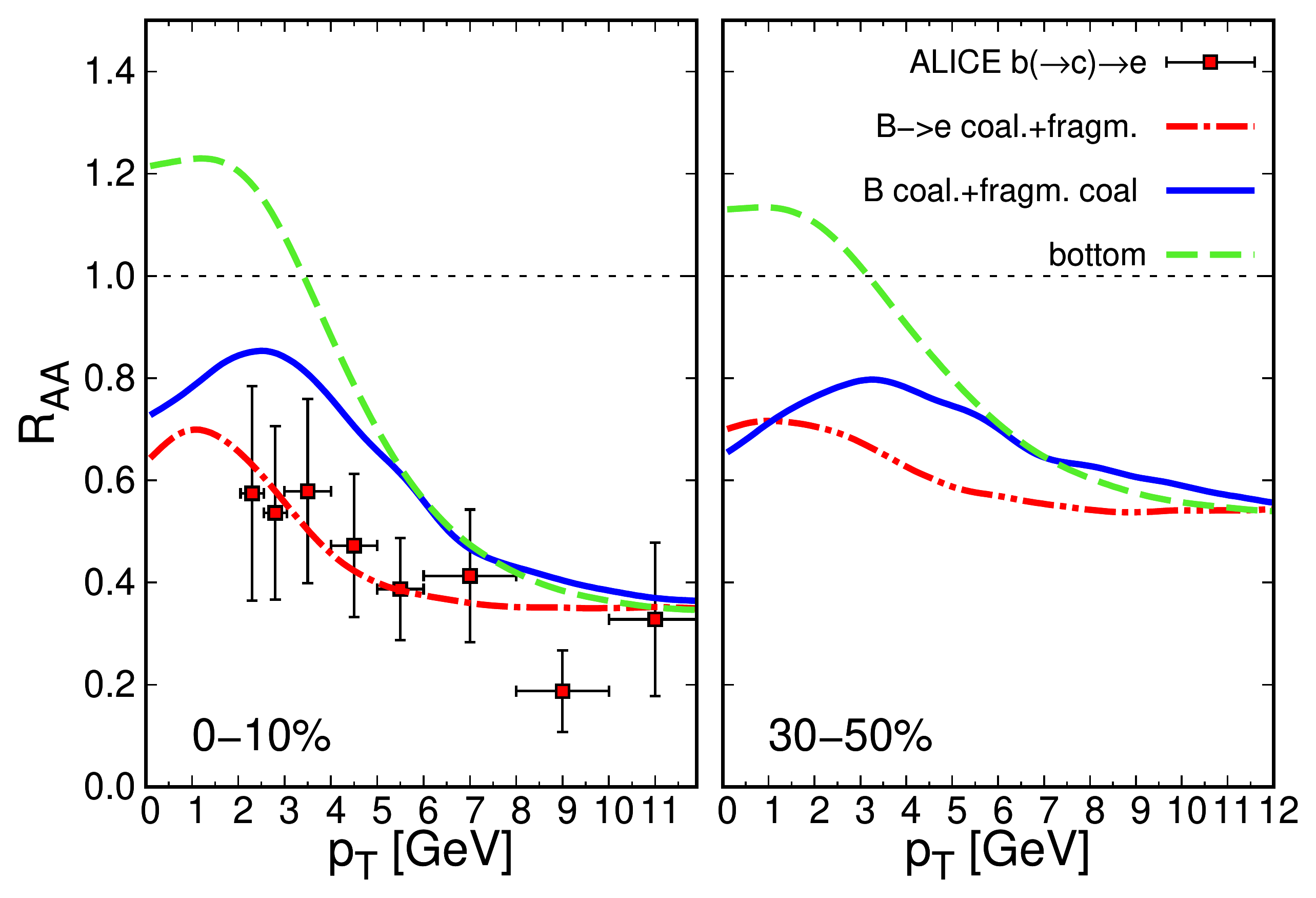}
			\caption{Nuclear modification factor $R_{AA}(p_T)$ for bottom quarks (green dashed line), $B$ mesons from coalescence plus fragmentation (blue solid line) and electrons from $B$ meson decay (red dot-dashed line) in $PbPb$ $\sqrt s= 5.02$ $TeV$ collisions at mid-rapidity and in $0-10\%$ (left panel) and $30-50\%$ (right panel) centrality class. The electrons nuclear modification factor at $0-10$ $\%$ is compared to the experimental measurements. Data taken from Ref. \cite{ALICE:2022iba}.}\label{Raa}
		\end{center}
	\end{figure}
	In order to evaluate the $R_{AA}(p_T)$, we have implemented in our code the decay channel $B(\rightarrow c)\rightarrow e$ taking into account the semi-leptonic decay matrix weighted by the different branching ratio of the decay.  In particular, we consider the semi-leptonic decay channel $\chi_b \rightarrow \chi_c + e + \nu_e $ characterized by a BR $\approx 10 \%$ and where $\chi_c$ describes the various species of D mesons.
	In Fig. \ref{Raa}, we show the nuclear modification factor $R_{AA}$ for bottom quark together with our prediction for $B$ mesons and electrons from $B$ meson decay in $PbPb$ $\sqrt s= 5.02$ $TeV$ collisions in both $0-10\%$ and $30 -50\%$ centrality classes. 
	The B mesons $R_{AA}$ has a behaviour similar to the one observed in the charm sector \cite{Scardina:2017ipo}. The hadronization via coalescence plus fragmentation gives,
	as expected, a shift of the peak to higher momenta which is smaller with respect to the one estimated with the same model for D mesons. This effect comes from the coalescence process that form B meson at a certain momentum combining bottom quarks with light quarks at lower momenta. 
	
	\begin{figure}[ht]
		\begin{center}
			\includegraphics[width=.45\textwidth]{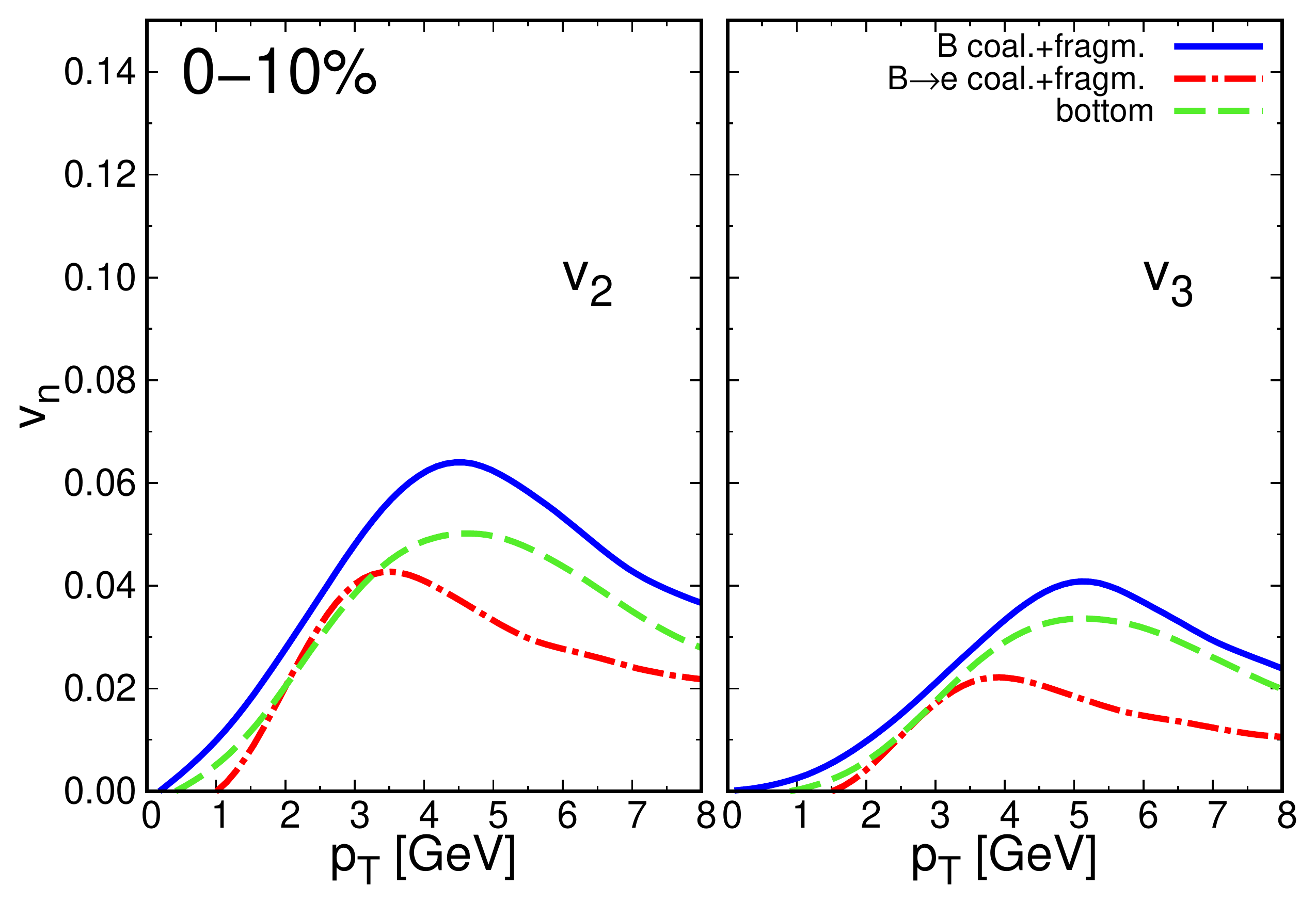}
			\caption{Elliptic flow $v_2(p_T)$ (left) and triangular flow $v_3(p_T)$ (right) for bottom quark, $B$ mesons and electrons from $B$ meson decay in $PbPb$ $\sqrt s= 5.02$ $TeV$ collisions in $0-10\%$ centrality class. Same legend as in Fig. \ref{Raa}.}\label{vn_010}
		\end{center}
	\end{figure}
	
\begin{figure}[ht]
		\begin{center}
			\includegraphics[width=.45\textwidth]{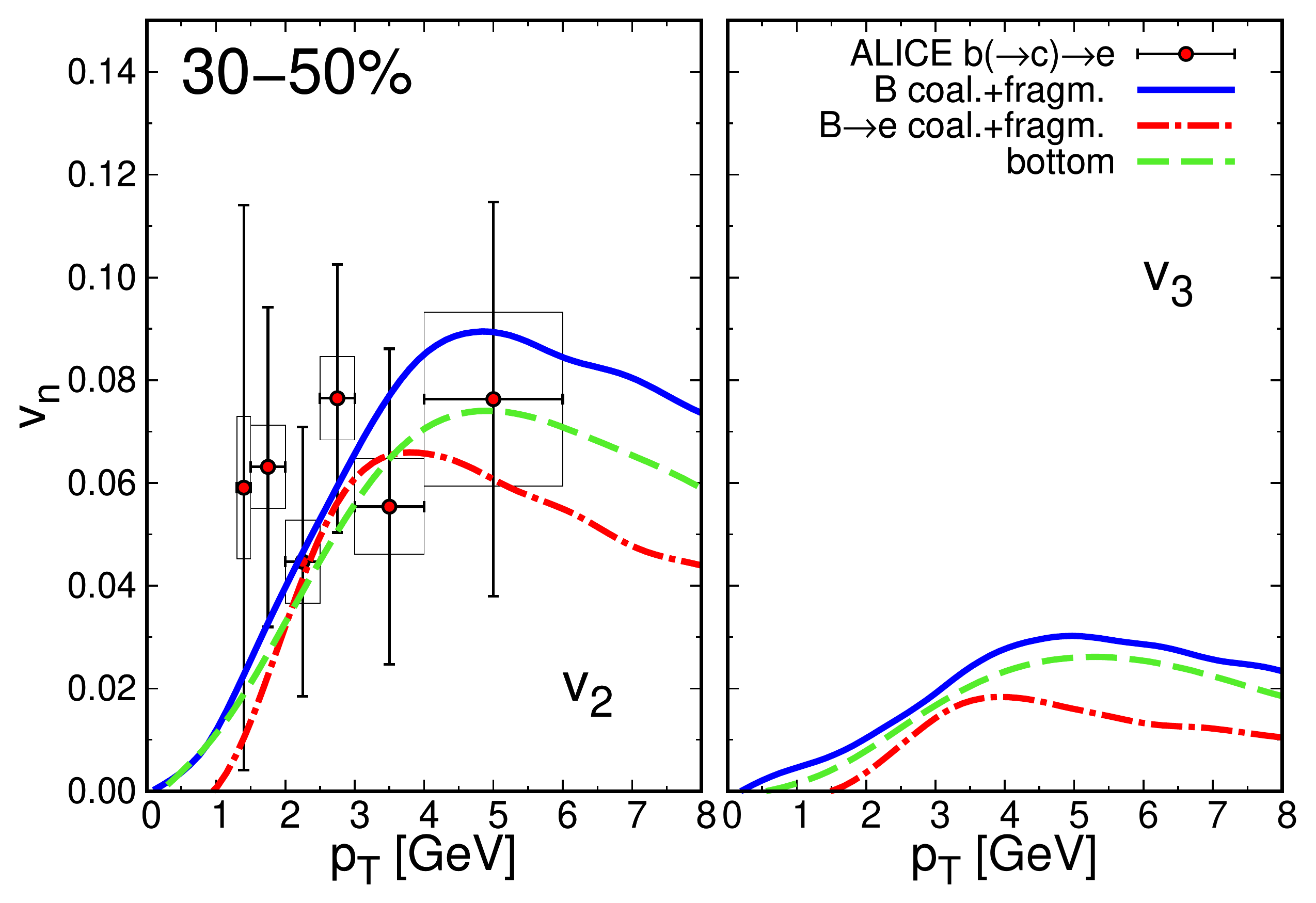}
			\caption{Elliptic flow $v_2(p_T)$ (left) and triangular flow $v_3(p_T)$ (right) for bottom, $B$ mesons and electrons from $B$ mesons decay in $PbPb$ $\sqrt s= 5.02$ $TeV$ collisions in $30-50\%$ centrality class. Same legend as in Fig. \ref{Raa}. The elliptic flow $v_2(p_T)$ of electrons is compared to the available experimental data from \cite{ALICE:2020hdw}.}\label{vn_3050}
		\end{center}
	\end{figure}

	We have also evaluated the $v_2(p_T)$ and $v_3(p_T)$ with two-particle correlation method for both $B$ meson and electrons from $B$ meson decay as shown in Fig. \ref{vn_010} and Fig. \ref{vn_3050} for both $0-10\%$ and $30-50\%$ centrality class respectively. We predict a non-zero elliptic flow $v_2$ and triangular flow $v_3$ for bottom quarks in both $0-10\%$ and $30-50\%$ centrality class. This suggests that bottom quarks take part in the collective motion in a way similar to what observed also for charm quarks \cite{Sambataro:2022sns,Plumari:2020eyx}, but with an efficiency of conversion of $\epsilon_{2}$ that is only about a $15\%$ smaller for $v_2$ in most central collisions and about $40\%$ smaller for $v_2$ at $30-50\%$ centrality. Regarding the conversion of $\epsilon_{3}$, we find an efficiency of conversion for bottom quark of about a $30\%$ smaller than charm quark for $v_3$ at both $0-10\%$ and $30-50\%$ centralities.
	We observe that moving from central to peripheral collision we get an enhancement of the elliptic flow as a consequence of the geometry of the overlapping region in more peripheral collisions which are characterized by a larger eccentricities $\epsilon_2$ ($\epsilon_2^{0-10\%} \simeq 0.13$ and $\epsilon_2^{30-50\%} \simeq 0.42$). On the other hand, our results show a comparable $v_3$ for $0-10\%$ and $30-50\%$ centrality class suggesting the triangular flow is generally related to the the event-by-event fluctuations of triangularity $\epsilon_3$ of the overlap region ($\epsilon_3^{0-10\%} \simeq 0.11$ and $\epsilon_3^{30-50\%} \simeq 0.21$). Similar results have been observed in light quark sector and recently in charm quark sector \cite{Sambataro:2022sns,Plumari:2015cfa,Niemi:2012aj}.
	
	Comparing the green dashed line with the blue solid lines we observe that the role of the hadronization is to give an enhancement of the anisotropic flows in both centralities that is about lost in the electrons from B decays where the $v_2$ and $v_3$ at $p_T \gtsim 3$ GeV in both centralities are on average at least a $25 - 30\%$ smaller than B mesons ones. Furthermore, our simulations show a good agreement between the $v_2$ and the available experimental data in $30 - 50\%$ centrality class from ALICE collaboration \cite{ALICE:2020hdw} suggesting that, despite the large mass, the coupling of bottom quarks to the bulk medium is strong enough to collectively drag them in the expanding fireball.
	Therefore, within the current data uncertainty our QPM approach is able to correctly predict available data, not only for D mesons \cite{Scardina:2017ipo}, but also for B mesons. This confirm QPM provide a reasonably good description of the HQ dynamics and in 
	particular the evolution with mass of the transport coefficient.
	
	\section{Spatial diffusion coefficient for charm and bottom}
	
	One usually compare the information on the HQ interaction in terms of the heavy quark spatial diffusion coefficient $D_s$, a quantity that, measuring the space dispersion per unit time, can be calculated also in lQCD and in the limit $p \rightarrow 0$. Furthermore, it can be related to the thermalization time of HQs:
	
	\begin{equation}\label{DS_final}
		\tau_{th}(p \rightarrow 0)= 
		\frac{M_{HQ}}{ 2\pi T^2} \, (2\pi T D_s)
	\end{equation}
	
\begin{figure}[ht]
		\begin{center}
			\includegraphics[width=.47\textwidth]{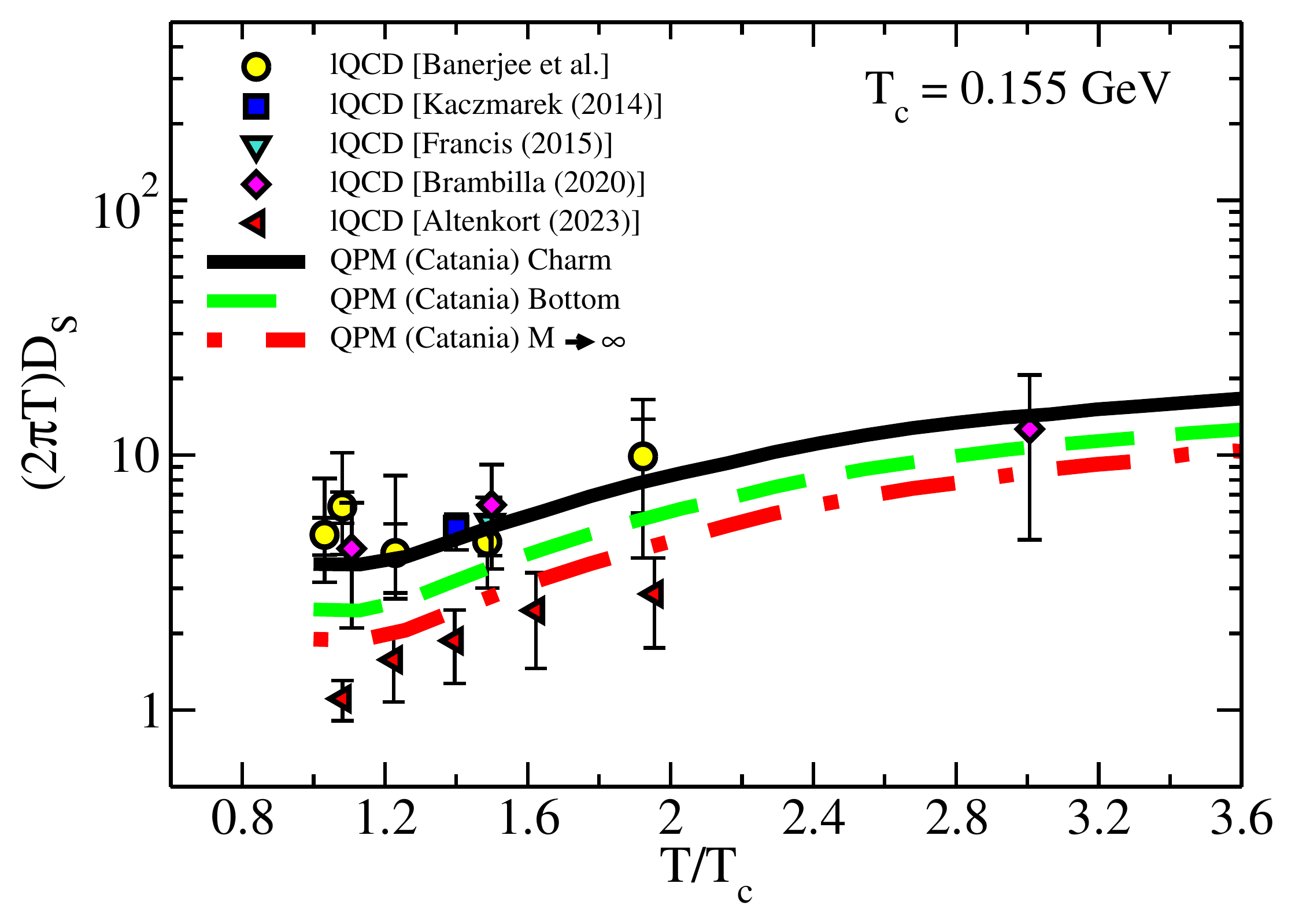}\caption{Spatial diffusion coefficient $D_s(T)$ for charm quark (solid black line) and bottom quark (dashed green line) compared to the lQCD expectations \cite{Banerjee:2011ra, Kaczmarek:2014jga, Francis:2015daa, Brambilla:2020siz, Altenkort:2023oms}. In the same panel we show $D_s(T)$ of charm quark opportunely scaled in order to reach the saturation scale of $M \rightarrow \infty$. For more details see Fig. \ref{massa_star}.}\label{Ds_scaled}
		\end{center}
	\end{figure}

	In our QPM approach, the scattering matrices for the interaction processes between the bulk and the heavy quarks are the same for charm and bottom, with the only difference coming from their masses ($M_c=1.4$ $GeV$ for charm quarks and $M_b=4.6$ $GeV$ for bottom quarks).
	In Fig. \ref{Ds_scaled}, the spatial diffusion coefficient $2\pi T\, D_s$ for both charm (black solid line) and bottom quarks (green dashed line) is shown in comparison to the available lQCD calculations. We can see from Fig.\ref{Ds_scaled} that $D_s$ of QPM for charm quark (black solid line) and
	lQCD data shows a good agreement within the current uncertainties, as already pointed out in several 
	Ref.s \cite{Dong:2019unq,Scardina:2017ipo,Greco:2017rro}. However, the lQCD data till 2020 are obtained in the infinite $M_Q$ limit and in a quenched medium, while the phenomenological QPM approach is for the finite HQ mass and in a medium including dynamical fermions. 
	Even if commonly discarded till now, it is important to study the $D_s$ dependence on heavy quark masses in our QPM approach in order to appropriately compare the results to the lQCD calculations.
	As mentioned above the charm mass, even if quite large with respect to  $\Lambda_{QCD}$, is yet comparable to the average momentum of the medium $\sim 2-3 \,T$ and the exchange momentum $q \sim \, g\, T$ \cite{Das:2013kea,Dong:2019unq}. Hence, it can be envisaged that its mass scale is not yet enough large to reach the limit where the thermalization time scales as $\tau_{th} \sim \frac{M_Q}{T}$$D_s$ (as usually assumed) and $D_s$ is mass independent. In fact, in Fig. \ref{Ds_scaled}, we can see that at $T_c$ the $D_s(T)$ in the QPM for the charm quark (black solid line) is about a $50\%$ larger than the bottom quark one (green dashed line) which is a significant breaking of the mass independence of $D_s(T)$ and implies a significant breaking of the $M/T$ scaling for the thermalization time $\tau_{th}$. The $2\pi T\,D_s(T)$ shown in Fig. \ref{Ds_scaled} for the charm quark correspond to an average thermalization time for low momenta of about $5\, \rm fm/c$  in the range of temperatures $1-2 \,T_c$. This would lead to estimate a bottom relaxation time $\tau_{th}(b) = (M_b/M_c) \tau_{th}(c) \sim 3.3 \, \tau_{th}(c) \sim 16.5 \,\rm fm/c$, according to the $\tau_{th} \sim M $ scaling in the large mass limit.  
	Instead the decrease of $D_s(T)$ with the mass of the HQ in the QPM implies a $\tau_{th}$ for bottom quark in the QPM is $\tau_{th}(b)\sim 11 \, \rm fm/c$, significantly smaller than the one extrapolated by a $M/T$ scaling of $\tau_{th}(c)$ and essentially only slightly larger than a factor of 2 w.r.t. the
	charm one.
	
\begin{figure}[ht]
		\begin{center}
			\includegraphics[width=.47\textwidth]{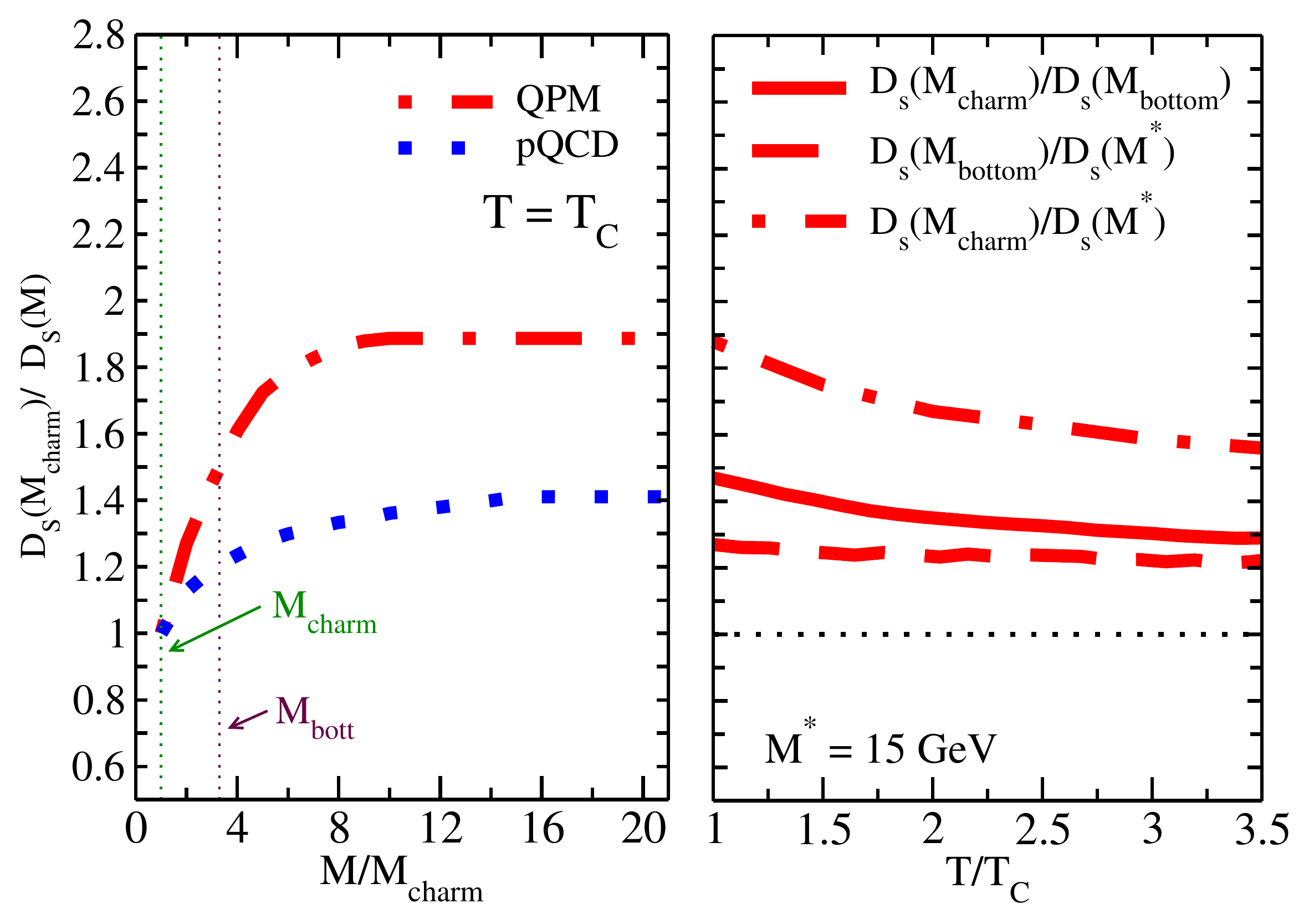}
			\caption{(left) Ratio $D_s(M_{charm})/D_s(M)$ as function of $M/M_C$ for both QPM and pQCD approach at $T=T_C$. (right) Ratio among spatial diffusion coefficient $D_s$ calculated within a QPM interaction for three different heavy-quark masses $M_{HQ}=1.4, 4.6, 15$ $GeV$ values.}\label{massa_star}
		\end{center}
	\end{figure}

	This aspect can be more clearly visualized in Fig.\ref{massa_star}, where we plot by red dot-dashed line the ratio between $D_s(M_{charm})$ and $D_s(M)$ as function of $M/M_{charm}$ both in pQCD and QPM approach at $T=T_C$. We can see that, in the region of charm and bottom masses, the $D_s$ is strongly mass dependent and reaches a saturation value only for masses $M_Q \sim 8 \, M_{charm} \gtsim 10 \, \rm GeV$ giving 
	$D_s(M_{charm})/ D_s(M \rightarrow \infty) \simeq \, 1.9$ .
	The effect, as can be expected, is stronger for the QPM that incorporates non-perturbative dynamics with respect to the case of pQCD where we find $D_s(M_{charm})/ D_s(M \rightarrow \infty) \simeq \, 1.4$. In the right panel of Fig. \ref{massa_star}, we show how the ratios $D_s(M_{charm})/D_s(M_{bottom})$ (solid red line), $D_s(M_{bottom})/D_s(M^\ast)$ (red dashed line) and $D_s(M_{charm})/D_s(M^\ast)$ (red dot-dashed line) evolve as function of temperature. Here
	the value $M^\ast=15$ $GeV$ ($M/M_{charm} > 10$) represents the mass of a fictitious super-heavy partner staying in the $M_Q \rightarrow \infty$.
	We note that we can assert that the bottom mass scale is quite close to the infinite mass limit with respect to the charm quark case with a discrepancy of only about a $20-25\%$; differently from the discrepancy between $D_s(M_{charm})$ and $D_s(M_{bottom})$ which is of about $50\%$ for $T=T_c=0.155$ $GeV$ and not smaller than $30\%$ at higher temperatures, while as said the $D_s(M_{charm})$  may lie
	up to about a factor of 1.5-2 above the $D_s(M \rightarrow \infty)$ as evaluated in lQCD.
	
	In Fig. \ref{Ds_scaled}, going back to the comparison to lQCD calculations, we have plot by red dot-dashed line the spatial diffusion coefficient in the $M_{HQ} \rightarrow \infty$ within the QPM approach. We observe that the $2\pi T \, D_s(T)$ in the large mass limit is quite close to the new lQCD data (red triangles \cite{Altenkort:2023oms}) which are obtained performing calculations in 2+1 flavours QCD with dynamical fermions differently from the other lQCD data obtained in quenched approximations. Therefore,
	while the QPM $D_s(T)$ for charm is quite close to the older lQCD data in quenched approximation,
	the $D_s(T)$ implied by QPM doing the appropriate comparison in the infinite mass limit, has a better agreement with lQCD simulations evaluated taking into account dynamical fermions that are the more pertinent one to compare to.

	\section{Conclusion}
	
	In this letter, we have studied the propagation of bottom quarks in the quark-gluon plasma (QGP) by means of an event-by-event Boltzmann transport approach. In particular, we have studied  within the QPM approach the $R_{AA}$ and $v_{2,3}$ at LHC energies of B mesons and electrons from semi-leptonic B meson decay in different centrality class selection. The study has been developed without any tuning, employing the same parameters of the model namely the coupling, as in previous studies on D mesons \cite{Scardina:2017ipo,Das:2015ana}. We find a good agreement with the available data on electrons for $R_{AA}(p_T)$ at $0-10\%$  and $v_{2}(p_T)$ at $30-50\%$ centrality, while for $R_{AA}(p_T)$ at $30-50\%$ and $v_{2,3}$ at $0-10\%$ and, in general directly for B mesons, data are not yet available.
	Our results suggest that bottom quark takes part in the collective expanding medium even if the large mass of bottom quark with respect to the charm one $M_B \sim 3.3 M_C$ leads to a mass ordering effect on the collective flows resulting in a smaller but still significant value for both $v_2$ and $v_3$ of B meson that mainly comes directly from a similarly large $v_{2,3}$ of the b quark. 
	
	Within kinetic theory in the $M/T \rightarrow \infty$ limit thermalization time should scale linearly with $M_{HQ}$, thus resulting in a $D_s(T)$ parameter which is a mass independent measure of the QCD interaction. 
	However, in the QPM approach the mass difference among charm and bottom
	leads to a $D_s(c)/D_s(b)$ ratio of about a factor of $1.5$ at $T \sim T_c$ decreasing slightly to $1.3$ at higher temperatures ($T\sim 3-4 \, \rm T_c$). This means that at the mass scale of charm quark the infinite mass limit used in lQCD is not yet reached; on the other hand, for the bottom mass scale, there is a discrepancy of only about a $20\%$ w.r.t. the infinite mass limit. 
	However, once the mass scale dependence is taken into account, the QPM approach appears to be in a satisfactory agreement with the most recent lQCD calculations that include dynamical fermions \cite{Altenkort:2023oms}, differently from previous lQCD data in quenched approximation.
	To our knowledge, this is the first time this aspect is explicitly discussed and quantified. However, it would be appropriate that the various modeling of the heavy quark dynamics, that aim to evaluate the $D_s(T)$ comparing to lQCD calculations, evaluate explicitly the mass dependence of $D_s(T)$ implicitly present in their approach with the aim of achieving a more pertinent and solid comparison to the new and upcoming lQCD calculations.
	Finally, we mention that a first estimate of thermalization time for bottom quark
	leads to values $\tau_{th}(b) \sim \, \rm 10-12 \,fm/c$ 
	which is about a factor of 2 larger than charm and so quite smaller that 3.3 as suggest by a simple 
	$M_{HQ}/T$ scaling.
	We however warn that such estimates of thermalization time through, Eq. \ref{DS_final}, 
	are in the $p \rightarrow 0$ while in the realistic case at finite momentum one should substitute $M_{HQ}$
	in Eq.\ref{DS_final} with the HQ kinetic energy. This, for a realistic charm quark distribution at LHC energy,
	means an increase of about a factor 2, while for bottom only a $30\%$. Hence to estimate the thermalization time of HQ in uRHICs
	directly from lQCD $2\pi T D_s(T)$ discarding both finite mass and momentum can lead to significantly underestimate the thermalization time of HQ, 
	in particular for charm quarks.

	\subsection*{Acknowledgments}
	V.G. acknowledges the funding from UniCT under ‘Linea di intervento 2’ (HQCDyn Grant). S.P. acknowledges the funding from UniCT under ‘Linea di intervento 3’ (HQsmall Grant). 
	We thank M. Ruggieri for useful discussion.
	
	
\end{document}